\documentclass[slac_one]{revtex4}
\usepackage{graphicx}
\usepackage{color}
\usepackage{fancyhdr}
\usepackage{xspace}
\def\met{\mbox{\ensuremath{\, \slash\kern-.67emE_{T}}}}

\def\lumi{$\cal{L}$}

\newcommand{\ttbar}{\ensuremath{t\bar{t}}}

\def\ifb{\mbox{fb$^{-1}$}}
\def\ipb{\mbox{pb$^{-1}$}}

\def\pT{\ensuremath{p_T}}

\def\ET{\ensuremath{E_T}}

\def\met{\mbox{\ensuremath{\, \slash\kern-.6emE_{T}}}}
\def\mpt{\mbox{\ensuremath{\, \slash\kern-.6emp_{T}}}}

\def\TeV{\ifmmode {\mathrm{\ Te\kern -0.1em V}}\else
                   \textrm{Te\kern -0.1em V}\fi\xspace}%
\def\GeV{\ifmmode {\mathrm{\ Ge\kern -0.1em V}}\else
                   \textrm{Ge\kern -0.1em V}\fi\xspace}%
\def\MeV{\ifmmode {\mathrm{\ Me\kern -0.1em V}}\else
                   \textrm{Me\kern -0.1em V}\fi\xspace}%
\def\keV{\ifmmode {\mathrm{\ ke\kern -0.1em V}}\else
                   \textrm{ke\kern -0.1em V}\fi\xspace}%
\def\eV{\ifmmode  {\mathrm{\ e\kern -0.1em V}}\else
                   \textrm{e\kern -0.1em V}\fi\xspace}%

\def\Vtb{\ensuremath{\vert V_{tb} \vert}\xspace}

\begin{document}

\title{Top quark physics expectations at the LHC} 

%

\author{Andrei~Gaponenko (for the ATLAS and CMS collaborations)}
\affiliation{LBNL, Berkeley, CA 94720, USA}
%

\begin{abstract}
The top quark will be produced copiously at the LHC.  This will make
both detailed physics studies and the use of top quark
decays for detector calibration possible.  This talk reviews plans and
prospects for top physics activities in the ATLAS and CMS experiments.
\end{abstract}

\maketitle

\thispagestyle{fancy}


\section{Introduction} 

ATLAS and CMS are general purpose detectors at the LHC, which will
provide proton-proton collisions with center of mass energy 14\TeV and
instantaneous luminosity up to $L=10^{34}\text{cm}^{-2}\text{s}^{-1}$.
Compared to the Tevatron, the production cross-section at the LHC is
expected to be two orders of magnitude larger for top quark pairs, but
only an order of magnitude larger for background events, giving large
improvements in both the available statistics and the signal to
background ratio.

%

The lepton+jets and dilepton decay channels of the \ttbar{} produce
high $p_t$ electrons, muons, jets, b-jets, and large missing
transverse momentum, exercising the whole detector system.
Observation of the \ttbar{} signal in LHC data will be an important
milestone in the physics commissioning of the experiments.

Both the ATLAS and CMS experiments have prepared multiple top physics
analyses and tested them on simulated data in anticipation of the
collider turn on.  This contribution mentions only a small subset of
top physics studies that have been performed.


\section{Establishing the signal}

\begin{figure*}[p]
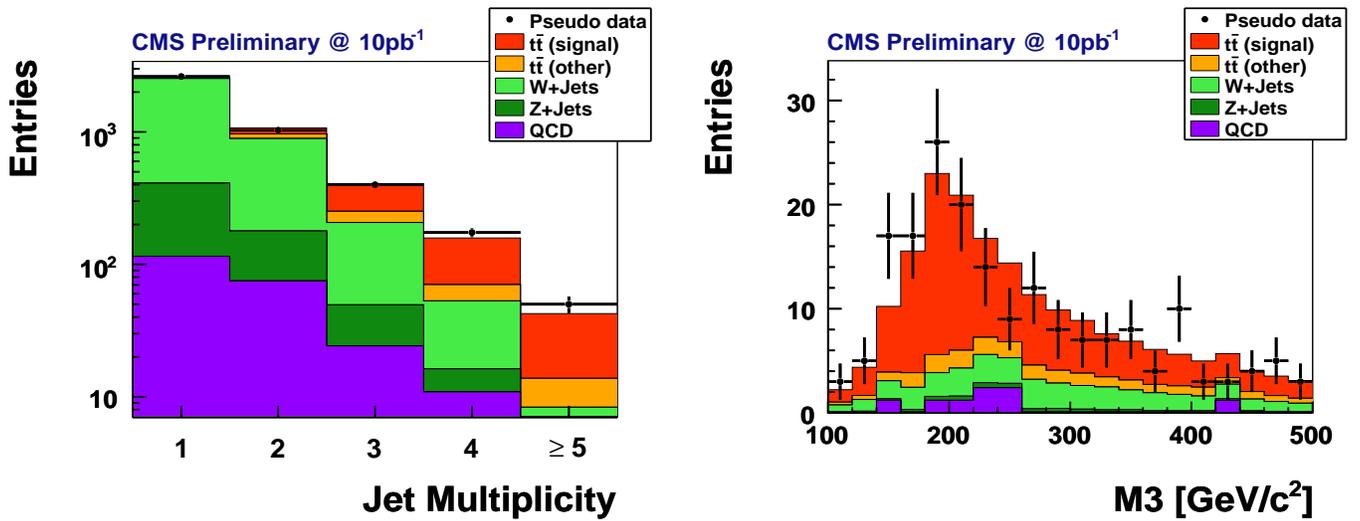

\includegraphics[width=0.48\textwidth,totalheight=0.29\textheight,keepaspectratio=0]{eps/cms-top-08-005-jem.epsi}
\hfill
\includegraphics[width=0.48\textwidth,totalheight=0.29\textheight,keepaspectratio=0]{eps/cms-top-08-005-mtop.epsi}
\caption{\label{fig:cmsXsecLJ}
  Jet multiplicity distribution (left) and the invariant mass of
  selected 3-jet combinations (right) in CMS for $10\ipb$ of simulated
  data in the  muon+jets  \ttbar{} analysis.}
\end{figure*}

The \ttbar{} production cross-section has been measured only at
Tevatron energies.  Measuring it at the LHC will provide a check on
theoretical predictions of the cross-section in the new energy regime.


A CMS study \cite{cms-lj-analysis} demonstrates that \ttbar{} signal
can be established in the muon+jets channel with as low as $10\ipb$ of
data.  The event selection requires exactly one isolated muon with
$\pT>30\GeV$ and $|\eta|<2.1$, and at least one jet with $\ET>65\GeV$.
For jets with $\ET>40\GeV$ and $|\eta|<2.4$ the left plot of
Figure~\ref{fig:cmsXsecLJ} shows the jet multiplicity distribution.
The final \ttbar{} selection is defined by requiring at least 4 jets.
The invariant mass of hadronically decaying top quark candidates,
which are formed by selecting 3-jet combinations with the highest
vector sum of the transverse momentum, is shown in
Figure~\ref{fig:cmsXsecLJ} on the right.  It has a clear peak near the
top quark mass.
%
%
B-tagging is not used in this study, and no requirement on missing
transverse energy is
made.  That makes this analysis suitable for the very early stage,
when these aspects of detector response are not well understood.


An ATLAS study \cite{atlas-csc-top-note} uses dilepton decay channels.
The event selection requires two isolated leptons ($ee$, $e\mu$ or
$\mu\mu$),  missing transverse energy $\met>35\GeV$, 
the direction of the missing transverse
momentum should be non-parallel to the muon, and $Z$ decays are
vetoed.  The selected events are used to produce a 2-dimensional
distribution of $\met$ vs the number of jets.  The ``data''
distribution is fit with a linear combination of Monte Carlo templates
for signal and background processes. 
It is found that a $5\sigma$ observation of the signal can be made
with $10\ipb$ of integrated luminosity, and with $100\ipb$ one can
expect the following precision on the cross-section measurement:
$
\Delta{\sigma}/\sigma =(4\text{(stat)}\pm4\text{(sys)}\pm2\text{(PDF)}\pm5\text{(lumi)})\%
$
Here systematics from QCD showering, ISR/FSR, jet energy scale,
trigger efficiency, electron identification efficiency and lepton
fakes are included in the ``(sys)'' number, 
and effects of the parton distribution
functions (PDFs) and luminosity uncertainties are shown separately.


\section{Use of top quarks for calibration}

\begin{figure*}[p]
\includegraphics[width=0.48\textwidth]{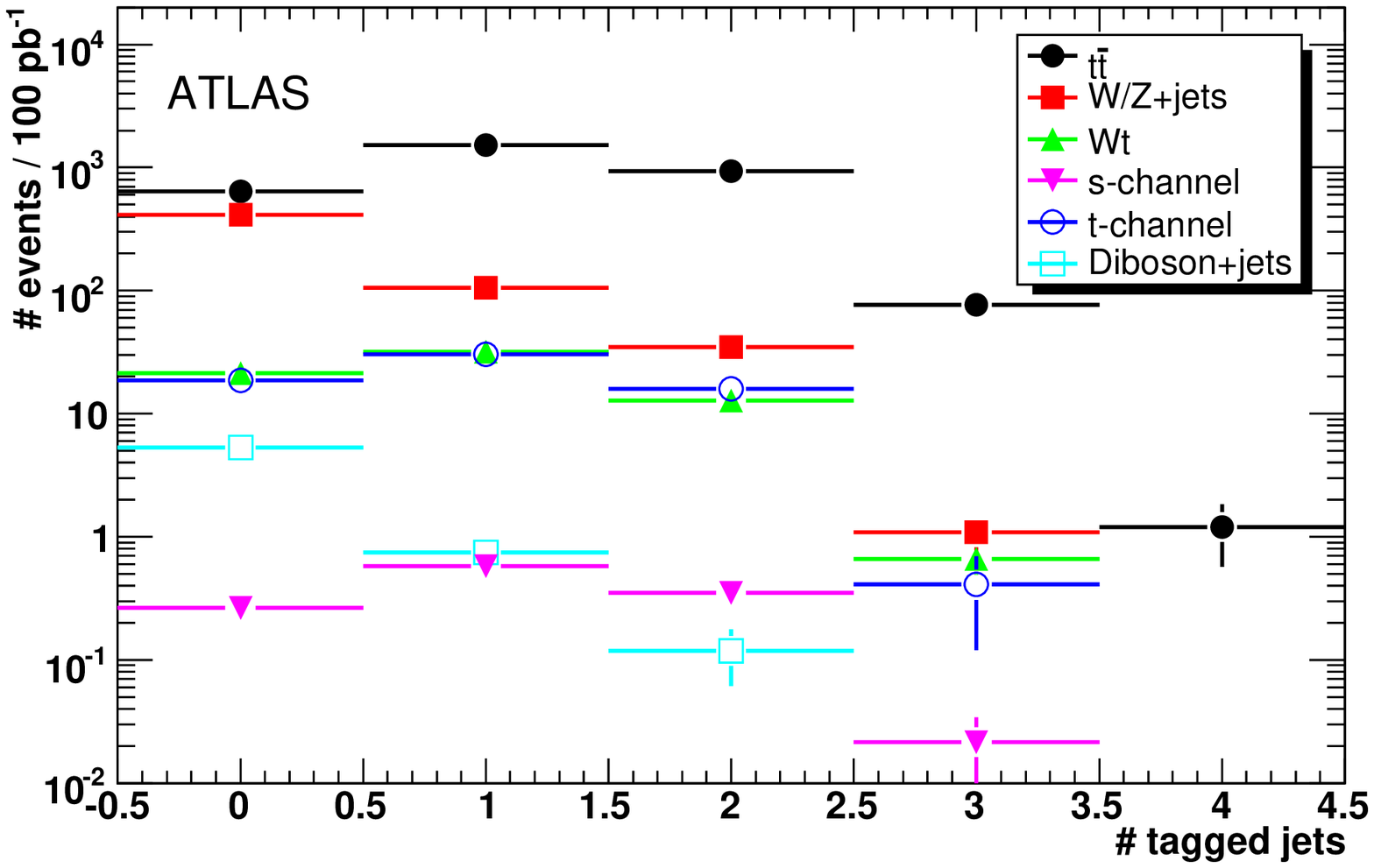}%
\hbox to0pt{\hss\raisebox{0.27\textwidth}{\colorbox{white}{\textsf{\textbf{ATLAS}}}}\hspace*{0.35\textwidth}}
\hfill
\includegraphics[width=0.48\textwidth,height=0.34\textwidth,keepaspectratio=0]{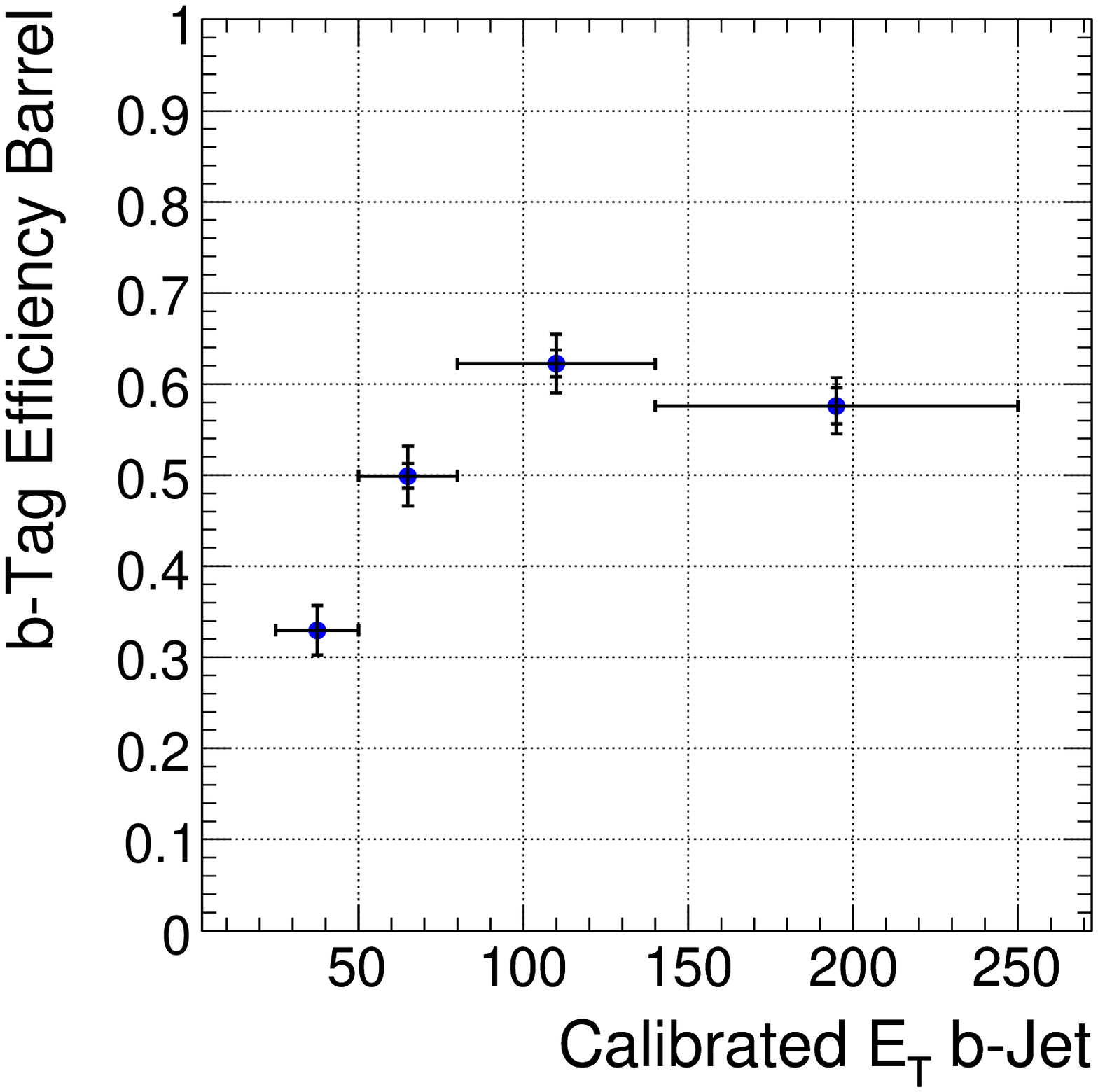}%
\hbox to0pt{\hss\raisebox{0.26\textwidth}{\textsf{\textbf{CMS}}}\hspace*{0.095\textwidth}}
\caption{\label{fig:b-tag-calib}Event yield in lepton+jets channel as
  function of the number of b-tagged jets in ATLAS for $100\ipb$
  (left). B-tagging efficiency in CMS barrel as function of jet
  $E_{T}$ for $1\ifb$ (right).}
\end{figure*}

The abundant production of top quarks at the LHC, and its well established
decay properties make it an excellent tool for calibrating the
detectors.  The higher multiplicities and transverse momenta of jets
in \ttbar{} decays, compared to other Standard Model processes, make
the calibration environment more similar to the one expected in many
New Physics searches.


The predominant decay mode of the top quark is $t\to{}Wb$
\cite{pdg08}.  One expects to find two b-jets in a
\ttbar{} event.  A way to extract the b-tagging efficiency $\epsilon_b$
from data is to count the number of b-tagged jets in a sample of
\ttbar{} events, and perform a Poisson likelihood fit to this
distribution.  An ATLAS study \cite{atlas-csc-top-note} demonstrates
that a relative precision of
${}\pm2.7\%\text{(stat)}\pm3.4\%\text{(sys)}$ on $\epsilon_b$ is
achievable with $100\ipb$ of data, for jets with $\ET>30\GeV$ at the
working point $\varepsilon_{b}=0.6$ in the lepton+jets
channel.  The distribution of the number of b-tagged jets in this
study is shown in Figure~\ref{fig:b-tag-calib} on the left.
Another method for measuring b-tagging efficiency with \ttbar{} events
is based on identification of a pure sample of b-jets by
reconstructing the \ttbar{} decay topology.  Unlike the tag counting
method described above, the topological selection method allows the
study of $\epsilon_b$ as a function of b-jet parameters.  A CMS study
\cite{cms-b-tagging} uses a likelihood ratio method to select a sample
of b-jets, and takes into account the impurities of the sample.  The
working point of the tagger is varied in the fit to minimize the
combined statistical and systematic uncertainty of the measurement.
The study combines lepton+jets and dileptonic \ttbar{} samples in
$1\ifb$ of simulated data.  The resulting $\epsilon_b$ in the barrel
region as function of b-jet $\ET$ is shown in
Figure~\ref{fig:b-tag-calib} on the right.  The expected precision on
$\epsilon_b$ with $1\ifb$ of data is ${}\pm6\%\text{(sys+stat)}$ in
the barrel region, and ${}\pm10\%\text{(sys+stat)}$ in the endcaps.


The well-known mass of the hadronically decaying $W$ in lepton+jets
\ttbar{} events can be used to calibrate jet energy scale.  An ATLAS
study \cite{atlas-csc-top-note} selects events with exactly one
isolated lepton with $\pT>20\GeV$, $\met>20\GeV$, at least 4 jets with
$\ET>40\GeV$, and exactly 2 b-tagged jets.  All light jet pair combinations
are used to produce an invariant mass distribution, which is fit
with a set of template histograms that differ in energy scale and
resolution parameters.  With $50\ipb$ of data the expected precision
on the light jet energy scale is 2\%, with systematic uncertainties
${}<0.5\%$.  With $1\ifb$ of data a 1\% precision should be
reachable.


\section{Physics studies with top}

\subsection{Top mass}

The mass of the top quark has been determined with a high precision at
the Tevatron.  The latest result that was shown in this conference
\cite{top-mass:2008vn},
$m_t=172.4\pm0.7\text{(stat)}\pm1.0\text{(sys)}$ , is systematics
dominated.  The large sample of \ttbar{} decays at the LHC will allows
the use of a tighter event selection to reduce systematic uncertainties and
improve the precision of the top mass measurement.  The Tevatron
measurements used sophisticated multivariate techniques in order to
extract maximum information from the limited data set. The first
measurements of top mass at the LHC will use simple cut-based
analyses, which may have different systematics, providing a powerful
cross-check of the result.

An example of such cut-based approach using lepton+jets channel is
presented by ATLAS in \cite{atlas-csc-top-note}.
Figure~\ref{fig:top-mass}, left, shows the distribution of reconstructed top
quark mass obtained with $1\ifb$ of simulated data, along with a fit
function.  The extracted top mass (for input
$m_{t,\text{true}}=175\GeV$) is
$
    m_t = 174.6 \pm 0.5\text{(stat)} 
    \pm 0.7/\%\text{(b-JES)}
    \pm 0.2/\%\text{(JES)}
    \pm 0.4\text{(ISR/FSR)}
    \pm 0.1\text{(b fragmentation)}
    \GeV
$.
With the expected precision on the light jet energy scale of 1\%, and on
the b-jet energy scale of between 1\% and 5\%, the precision on the top quark
mass will be between $1\GeV$ and $3.5\GeV$ for $1\ifb$ of data.

%

\subsection{Single top studies}


The sensitivity of single top cross-section to beyond the Standard
Model processes is studied in \cite{Tait:2000sh}.  The paper concludes
that the $Wt$ production channel is not affected by most New Physics
models and can provide a model-independent measurement of $\Vtb$,
while s- and t-channels receive different contributions from different
New Physics models.  Measurements of individual s- and t- channel
cross-sections may provide a handle to determine the kind of New
Physics contribution if one is found.

At the LHC, the expected single top channel cross-sections are
$\sigma_t\approx250$~pb, $\sigma_{Wt}\approx66$~pb, and
$\sigma_s\approx11$~pb.  Measuring the cross-sections is complicated
by large backgrounds from \ttbar{} and $W$+jets processes.
Some studies in \cite{atlas-csc-top-note} that use cut-based, boosted
decision tree (BDT), and likelihood approaches are summarized in
Table~\ref{t:single-top}.  For example, the BDT analysis in the t-channel
will allow the measurement of $\Vtb$ with relative precision
${}\pm11\%\text{(stat+sys)}\pm4\%\text{(theor)}$.

\begin{table}
\caption{\label{t:single-top}Expected precision on single top
  production cross-section in different channels.}
\begin{tabular}{r|c|c|r|l}
 & method & S/B & \hbox{}\hfill\lumi\hfill\hbox{} & xsec precision\\
\hline
$t$-chan & cuts & 0.37 &  $1\ifb$ & ${}\pm5\%\pm45\%$ \\
$t$-chan & BDT & 1.3 & $1\ifb$ & ${}\pm6\%\pm22\%$ \\
$Wt$ & BDT & 0.35 & $10\ifb$ & ${}\pm20\%$ \\
s-chan & likelihood & 0.19 & $30\ifb$ & $3\sigma$ evidence
\end{tabular}
\end{table}



\subsection{\ttbar{} resonances}

The large value of the top mass, which is close to the electroweak
symmetry breaking (EWSB) scale,  may point to a special role
that the top quark plays in the symmetry breaking.  New resonances or
gauge bosons strongly coupled to the top quark, such as in
technicolor, topcolor, or other strong EWSB models, or in models with
extra dimensions, could be manifest in the invariant mass distribution
of top quark pairs.

The experimental challenges are \ttbar{} reconstruction efficiency,
which drops as the resonance mass increases because decay products of
a boosted top quark start to overlap, and the resolution of the
reconstructed \ttbar{} mass.  
%
%
A generic narrow
resonance at $700\GeV$ with $\sigma\times\text{Br}(Z'\to{}\ttbar)=11\text{~pb}$ can
be discovered with a $5\sigma$ significance after $1\ifb$ of data
taking.  The discovery potential for a Kaluza-Klein gluon resonance is
shown in Figure~\ref{fig:ttbar-resonances} on the right.  With $1\ifb$
of data a $1.5\TeV$ resonance will be discovered with a $5\sigma$
significance.

\begin{figure*}[p]
\includegraphics[width=0.48\textwidth]{eps/t9.geometric-mt-fit.epsi}%
\hbox to0pt{\hss\raisebox{0.09\textwidth}{\textsf{\textbf{ATLAS}}}\hspace*{0.06\textwidth}}
\hfill
\includegraphics[width=0.48\textwidth]{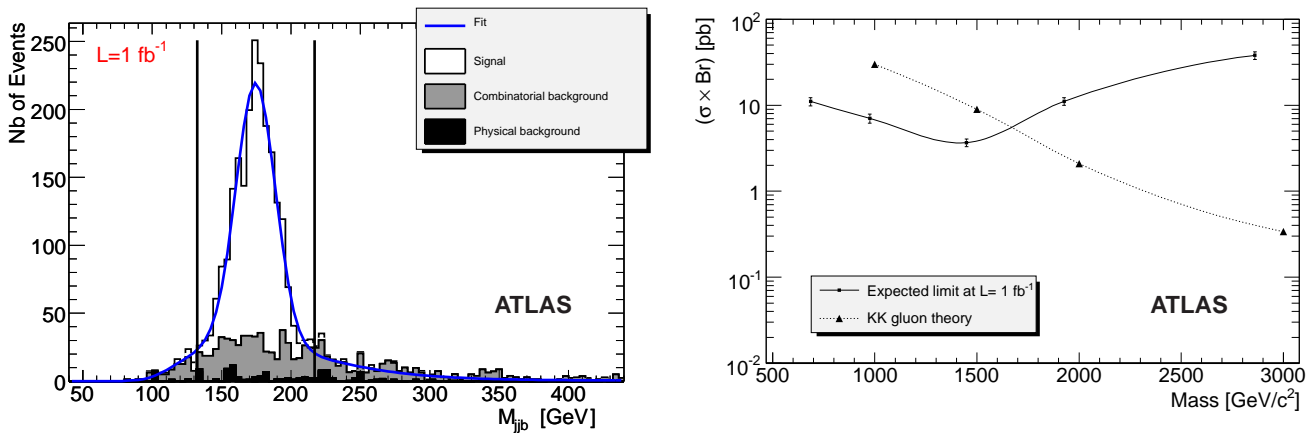}%
\hbox
to0pt{\hss\raisebox{0.2920\textwidth}{\colorbox{white}{\makebox[0.15\textwidth]{\vrule
        width0pt height0.0015\textwidth\hfill}}}\hspace*{0.04\textwidth}}
\hbox
to0pt{\hss\raisebox{0.285\textwidth}{\colorbox{white}{\makebox[0.008\textwidth]{\vrule
        width0pt height0.0025\textwidth\hfill}}}\hspace*{0.04\textwidth}}
\hbox to0pt{\hss\raisebox{0.09\textwidth}{\textsf{\textbf{ATLAS}}}\hspace*{0.09\textwidth}}
\caption{\label{fig:top-mass}\label{fig:ttbar-resonances}
  Left: reconstructed mass in the ATLAS cut-based
  top mass analysis with $1\ifb$ of simulated data.
  Right:   theoretical cross-section of
  Kaluza-Klein gluon resonance and ATLAS $5\sigma$ discovery potential.}
\end{figure*}

%

\begin{acknowledgments}

I would like to thank everyone in ATLAS and CMS collaborations whose
work contributed to this talk.  I am particularly grateful to Beate
Heinemann for her valuable comments and suggestions.

%

This work was supported by the Director, Office of Science, Office of
High Energy Physics, of the U.S. Department of Energy under Contract
No. DE-AC02-05CH11231.

\end{acknowledgments}


\end{document}